\documentclass[fleqn,usenatbib]{mnras}

\usepackage{newtxtext,newtxmath}

\usepackage[T1]{fontenc}
\usepackage{ae,aecompl}

\usepackage{graphicx}   
\usepackage{amsmath}    
\usepackage{amssymb}    

\usepackage{breakurl}

\newcommand{\kms}{~km s$^{-1}$~}
\newcommand{\AG}{AG~Peg~}
\newcommand{\AGE}{AG~Peg}
\newcommand{\dotM}{~M$_{\odot}$~yr$^{-1}$~}
\newcommand{\XMM}{{\it XMM-Newton~}}
\newcommand{\XMME}{{\it XMM-Newton}}

\newcommand{\Swift}{{\it Swift~}}
\newcommand{\SwiftE}{{\it Swift}}
\newcommand{\Rosat}{{\it ROSAT~}}
\newcommand{\RosatE}{{\it ROSAT}}
\newcommand{\xspec}{{\sc xspec~}}
\newcommand{\xspecE}{{\sc xspec}}

\title[X-rays from \AGE]
{An \XMM observation of the symbiotic star \AGE:
the X-ray emission after the end of its 2015 outburst}

\author[S.A.Zhekov and T.V.Tomov]{Svetozar A. Zhekov$^1$\thanks{
E-mail: szhekov@astro.bas.bg; toma.tomov@astri.umk.pl.} 
and Toma V. Tomov$^2$\\
$^1$Institute of Astronomy and National Astronomical Observatory
(Bulgarian Academy of Sciences),\\
72 Tsarigradsko Chaussee Blvd., Sofia 1784, Bulgaria\\
$^2$Centre for Astronomy, Faculty of Physics, Astronomy and
Informatics, Nicolaus Copernicus University, \\
Grudziadzka 5, 87-100 Torun, Poland
}

\date{}

\pubyear{2018}

\begin{document}
\label{firstpage}
\pagerange{\pageref{firstpage}--\pageref{lastpage}}
\maketitle

\begin{abstract}
We present an analysis of the \XMM observation of the symbiotic 
star \AGE, obtained after the end of its 2015 outburst.
The X-ray emission of \AG is soft and of thermal origin. \AG is 
an X-ray source of class $\beta$ of the X-ray sources amongst the 
symbiotic stars, whose X-ray spectrum is well matched by a
two-temperature optically-thin plasma emission ($kT_1 \sim 0.14$~keV 
and $kT_2 \sim 0.66$~keV). The X-ray emission of the class $\beta$
sources is believed to originate from colliding stellar winds (CSW) in
binary system. If we adopt the CSW picture, the theoretical
CSW spectra match well the observed properties of the \XMM spectra of
\AGE.  However, we need a solid evidence that a massive-enough hot-star
wind is present in the post-outburst state of \AG to proof the 
validity of the CSW picture for this symbiotic binary.
No short-term X-ray variability is detected while the UV emission of 
\AG shows stochastic variability (flickering) on time-scales of
minutes and hours.

\end{abstract}

\begin{keywords}
shock waves -- stars: individual: \AG -- stars: binaries: symbiotic --
X-rays: stars.
\end{keywords}



\section{Introduction}
\AG (HD 207757) is a symbiotic nova whose 
first recorded
outburst occurred in the mid-19th century. 
This is likely the slowest classical nova eruption amongst
recorded until present day (e.g., \citealt{kenyon_93}; 
\citealt*{kenyon_01} and references therein).
During the time of declining brightness (lasting for many decades), 
\AG has passed through 
various spectral phases, as signs of new active phase were recently
reported both in the optical \citep{munari_13} and in X-rays 
(\citealt{nunez_13}; \citealt{luna_15}; \citealt{ramsay_15}). 
In fact, 2015 marked an outburst\footnote{Throughout this text, the
term {\it outburst} will be used to denote the 2015 activity of \AGE.}
of \AG with two brightness maxima in
the optical and  the corresponding analysis indicated 
that \AG became a member of the classical symbiotic stars group
(\citealt{tomov_16}; \citealt{ramsay_16}; \citealt{skopal_17}).

In X-rays, \AG was observed numerous times. 
It was first detected with \Rosat \citep{murset_95}
years before its active phase in 2015. The X-ray emission of this
pre-outburst state was soft likely originating from an optically thin
plasma with a temperature of a few 10$^6$~K, which marked \AG as a 
member of the class $\beta$ of the X-ray sources amongst symbiotic 
stars \citep{murset_97}. We note that 
the X-ray emission of these sources is believed to
originate from colliding stellar winds (CSW) in binary system (e.g.,
\citealt{murset_97}; \citealt{luna_13}). 

The \Swift observations during the 2015 outburst revealed considerable 
X-ray variability on time-scales of days that resembles the
characteristics of the flicker noise (flickering), typical for
accretion processes in astrophysical objects \citep{zht_16}.
But, the X-rays might originated in shocks from the ejecta
\citep{ramsay_16}.
Such a diversity of properties of the X-ray emission from \AG
justifies the need of follow-up observations and such were proposed.

In this paper, we report results from the \XMM observation of \AGE,
carried out after the end of its active phase in 2015.
In Section~\ref{sec:data}, we review the observational data.
In Section~\ref{sec:results}, we present results from analysis of the
X-ray emission of \AGE. In Section~\ref{sec:discussion}, we discuss 
our results and Section~\ref{sec:conclusions} present our conclusions.

\section{Observations and data reduction}
\label{sec:data}

\begin{figure}
\begin{center}
\includegraphics[width=\columnwidth]{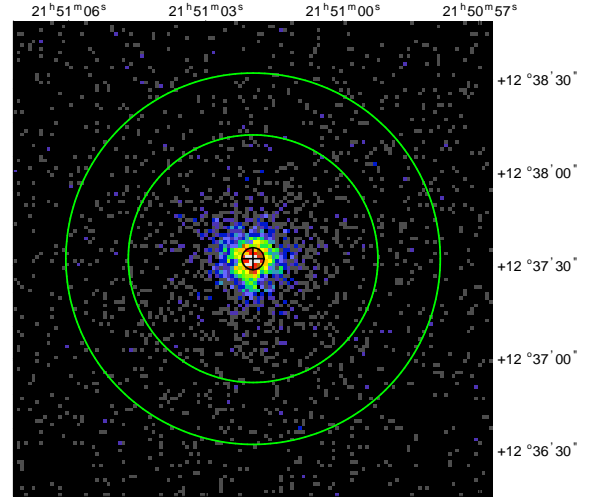}
\end{center}
\caption{The raw EPIC-MOS1 image of \AG in the (0.2 - 10 keV) energy
band with the spectral extraction regions. The source spectrum was
extracted from the central circle, while the background spectrum was
extracted from adjacent annulus. The circled plus sign gives the
optical position of \AG (SIMBAD).
}
\label{fig:image}
\end{figure}

\AG was observed with \XMM on 2017 Nov 16 (Observation ID 0800420101)  
with a nominal exposure of $\sim 30.6$ ks. The source was clearly 
detected in X-rays (see Fig.~\ref{fig:image}) and good quality data
were acquired with the European Photon Imaging Camera 
(EPIC) having one pn and two MOS detectors. We made use of the
data from the two instruments of the Reflected Grating Spectrometer (RGS)
as well, but we note that these data had quite limited photon statistics.
The \XMM Optical Monitor (OM)\footnote{for EPIC, RGS
and OM see \S~3.3, 3.4 and 3.5 in the 
\XMM Users Handbook,
\url{https://xmm-tools.cosmos.esa.int/external/xmm_user_support/documentation/uhb/}
}
telescope allows for obtaining optical/UV data on a given target 
simultaneously with its X-ray data. Thus, six UV exposures in UVM2
filter were obtained that provided good quality UV light curves of
\AGE.

{\it X-rays.}
For the data reduction, we made use of the \XMM 
{\sc sas}\footnote{Science Analysis Software, 
\url{https://xmm-tools.cosmos.esa.int/external/xmm_user_support/documentation/sas\_usg/USG/}}
16.1.0 data analysis software.
The {\sc sas} pipeline processing scripts emproc, epproc and rgsproc 
were executed to incorporate the most recent calibration files (as of 
2017 December 20). The data were then filtered for high X-ray 
background following the instructions in the {\sc sas} documentation. 
The corresponding {\sc sas} procedures were adopted to generate the
response matrix files and ancillary response files for 
each spectrum. The MOS spectrum in our analysis is the sum of the
spectra from the two MOS detectors. 
The extracted EPIC spectra (0.2 - 10 keV) of \AG had
$\sim 4786 $~source counts in the 10.5-ks pn effective exposure and 
$\sim 5272$~source counts in the 28.2-ks MOS effective exposure.
The high-resolution spectra had only $\sim 300$ (RGS1) and $\sim 500$ 
(RGS2) source counts in the 29.4-ks effective exposure.
Also, we constructed the pn and MOS1,2 background-subtracted light 
curves of \AGE. 

{\it UV.}
We used the pipeline background-subtracted light 
curves of \AG in the UVM2 filter (effective wavelength and width of 
2310 \AA ~and 480 \AA, respectively) in units of counts s$^{-1}$ which
can be presented also in magnitudes using the zeropoint for
the UVM2 filter. 

For the spectral analysis, we used standard as well as custom
models in version 12.9.1 of \xspec \citep{Arnaud96}.

\section{Results}
\label{sec:results}

\begin{figure*}
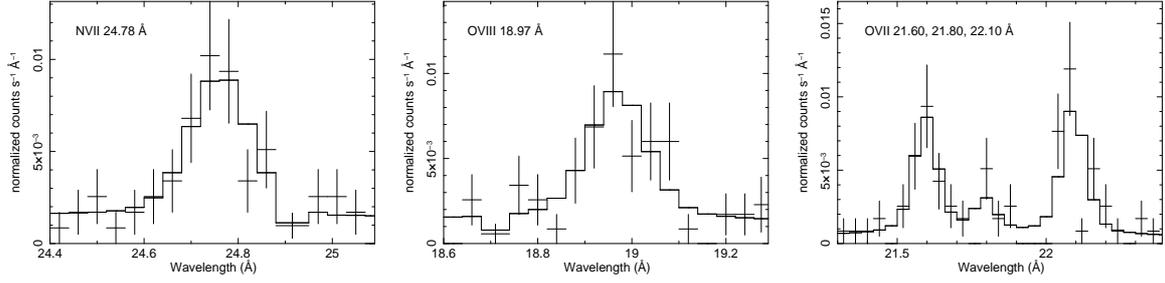

\begin{center}
\includegraphics[width=1.43in, height=2.0in, angle=-90]{fig2a.eps}
\includegraphics[width=1.43in, height=2.0in, angle=-90]{fig2b.eps}
\includegraphics[width=1.43in, height=2.0in, angle=-90]{fig2c.eps}
\end{center}
\caption{Line profiles of the NVII and OVIII  L$_\alpha$ emission
lines (H-like doublets) and of the OVII K$_\alpha$ emission line
(He-like triplet) in the RGS2 and RGS1 spectrum of \AGE, respectively.
The model fits are shown with the solid line. Spectra are slightly
re-binned for presentation.
}
\label{fig:lines}
\end{figure*}

\subsection{An overview of the X-ray emission}
\label{sec:xray_overview}
The \XMM observation of \AG provided data of much better quality 
(e.g., photon statistics, spectral resolution) than that of
the previous X-ray observations (\RosatE, \SwiftE) of this object. 
This allows us to better constrain the properties of its X-ray 
emission.

An important piece of information comes from the \XMM dispersed
spectra of \AGE. Namely, emission lines of various ionic species are
detected. Figure~\ref{fig:lines} shows RGS1 and RGS2 data for the
strongest H-like doublets (NVII and OVIII) and the OVII He-like 
triplet. Although the photon statistics of these spectra is
not high, the line detection is reliable (at a $4-5\sigma$ level): 
$46 \pm 9$ (NVII), $46 \pm 9$ (OVIII) and $40 \pm 11$ (OVII) source
counts. Other lines as the FeXVII 15.01, 15.21 \AA ~and 16.78, 17.05,
17.10 \AA ~are also detected but at lower confidence level.

We fitted the H-like doublets with a sum of two Gaussians and a
constant continuum. The centres of the doublet components were held
fixed according to the AtomDB database (Atomic Data for
Astrophysicists)\footnote{For AtomDB see \url{http://www.atomdb.org/}}
and the components shared the same line width and line shift. The
component intensity ratio was fixed at its atomic data value.
The OVII He-like triplet was considered in a similar manner but its
component ratios were free parameters of the fit.
For these fits, we made use of the implementation of the Cash statistic 
\citep{cash_79} in \xspecE, which gives robust results for model fits
of data with low photon counts.

The basic results from these fits are as follows. The strong emission
lines show no appreciable line shifts (all are consistent with
zero-shifts) and the line widths (full width at half maximum, FWHM)
are of the order of 700-1500 \kms. But, it should be kept in mind that 
the dispersed spectra of \AG have quite limited photon statistics, so, 
these fit results could be affected by that. Nevertheless, we note that
the forbidden line (at 22.1 \AA) in the OVII He-like triplet is 
{\it not} suppressed as seen from Fig.~\ref{fig:lines} and its ratio 
to the inter-combination line (at 21.8 \AA) is $4.07\pm2.60$. We 
will return to this result in Section~\ref{sec:discussion}.

However, the most important result from the dispersed spectra of \AG
is that emission lines are indeed detected. This is a solid evidence 
that the X-ray emission of \AG in 2017 is of {\it thermal origin}, 
that is it originates in optically thin thermal plasmas.

\begin{figure}
\begin{center}
\includegraphics[width=2.0in, height=2.8in, angle=-90] {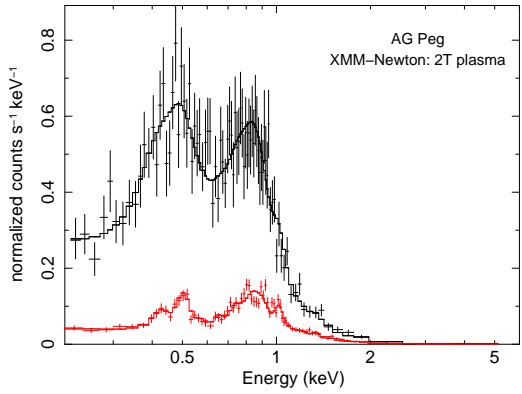}
\end{center}
\caption{The background-subtracted EPIC spectra of \AG overlaid with
the two-temperature optically thin plasma  model.
The pn (upper curve) and MOS (lower curve) spectra are
shown in black and red colour, respectively.
The spectra were re-binned to have a minimum of 50 counts per bin.
}
\label{fig:spec_2T}
\end{figure}

Since the X-ray source is thermal and almost all of its emission is 
at energies below 2 keV (see Fig.~\ref{fig:spec_2T} and 
Section~\ref{sec:xray_spec}), it is 
conclusive that \AG in 2017 (\XMM observation) is of the class $\beta$ 
of the X-ray sources amongst the symbiotic stars 
(for the X-ray source classification see \citealt{murset_97} and
\citealt{luna_13}).

\begin{table}
\caption{Global Spectral Model Results
\label{tab:fits}}
\begin{tabular}{ll}
\hline
\multicolumn{1}{c}{Parameter} & \multicolumn{1}{c}{2T vapec}  \\
\hline
$\chi^2$/dof  &  168/151 \\
N$_{\mathrm{H}}$ (10$^{21}$ cm$^{-2}$)  & 0.77$^{+0.22}_{-0.17}$ \\
kT$_1$ (keV) &  0.14$^{+0.01}_{-0.01}$ \\
kT$_2$ (keV) &  0.66$^{+0.02}_{-0.02}$ \\
EM$_1$ ($10^{54}$~cm$^{-3}$) &  1.78$^{+0.77}_{-0.47}$ \\
EM$_2$ ($10^{54}$~cm$^{-3}$) &  3.32$^{+0.44}_{-0.39}$ \\
O   &  0.67$^{+0.11}_{-0.11}$  \\
Ne  &  0.86$^{+0.22}_{-0.19}$  \\
Mg  &  0.39$^{+0.13}_{-0.11}$  \\
Si  &  0.20$^{+0.11}_{-0.10}$  \\
Fe  &  0.23$^{+0.04}_{-0.03}$  \\
F$_{X}$ ($10^{-13}$ ergs cm$^{-2}$ s$^{-1}$)  &  6.33 (11.8) \\
F$_{X,hot}$ ($10^{-13}$ ergs cm$^{-2}$ s$^{-1}$)  &  4.73 (7.33) \\
L$_{X}$ ($10^{31}$ ergs s$^{-1}$)  &  9.06 \\
\hline

\end{tabular}

{\it Notes.}
Results from  simultaneous fits to the EPIC
spectra of \AGE.
Tabulated quantities are the neutral H absorption column density
(N$_{\mathrm{H}}$), plasma temperature (kT),
emission measure ($\mbox{EM} = \int n_e n_{H} dV $),
the absorbed X-ray flux (F$_X$; 0.2 - 8 keV) followed
in parentheses by the unabsorbed value
(F$_{X,hot}$~ denotes the higher-temperature component, kT$_2$),
the X-ray luminosity L$_{X}$ (0.2 - 8 keV).
The EM and L$_{X}$ values assume
a reference distance of 800 pc 
(\citealt{kenyon_93}; \citealt{kenyon_01}),
The derived abundances are with
respect to the solar abundances \citep{ag_89}. 
Errors are the $1\sigma$ values from the fits.

\end{table}

\subsection{Global Spectral Models}
\label{sec:xray_spec}

To derive some physical properties of the thermal plasma responsible 
for the X-ray emission from \AGE, we fitted the observed spectra 
(pn, MOS) with  models of absorbed optically thin plasma emission. 
The pn and MOS spectra were fitted simultaneously sharing identical
model parameters.
In the spectral fits, the H, He, C and N abundances had the values
derived by
\citet{schmutz_96}: H : He : C : N = 1 : 0.1 : $10^{-5}$ : $10^{-3}$ 
by number (H $= 1$, He $= 1.02$, C = $0.03$, N $= 8.93$ with respect 
to the solar abundances; \citealt{ag_89}).
To improve the quality of the fits, the abundances of some elements 
(O, Ne, Mg, Si and Fe) were allowed to vary.

We performed one- and two-temperature plasma model fits: the latter
provides a better match to the observed spectra 
(the goodness of the fit is 0.0008 for the former and 0.17 for the
latter).
Table~\ref{tab:fits} and Fig.~\ref{fig:spec_2T} present the
corresponding results from the two-temperature fit to the undispersed 
\XMM spectra of \AGE.

As seen from Fig.~\ref{fig:spec_2T}, the X-ray emission from \AG is
soft: almost all the photons have energy $\leq 2$~keV. In terms of
observed flux, its part in the 0.2 - 2 keV energy range comprises 96
per cent of the total observed flux in the 0.2 - 8 keV range
(for the latter see Table~\ref{tab:fits}).

Such a soft emission is naturally produced by thermal plasma with
relatively low temperature as derived in the spectral fits. We note
that the two-component plasma model adopted here, which is a good
representation of the observed spectrum, in fact indicates a
distribution of plasma in \AGE. Namely, the X-ray emitting region of 
\AG is likely temperature stratified 
and the hotter the plasma the higher its emission measure.

The derived value of the hydrogen absorption column density is
equivalent to a 
range of optical extinction A$_{\mathrm{V}} = 0.35 - 0.47$ mag. 
The range corresponds to the conversion that is used:
N$_{\mathrm{H}} = 2.22\times10^{21}$A$_{\mathrm{V}}$~cm$^{-2}$ \citep{go_75}.
and 
N$_{\mathrm{H}} = (1.6-1.7)\times10^{21}$A$_{\mathrm{V}}$~cm$^{-2}$
(\citealt{vuong_03}, \citealt{getman_05}).
Thus, we find no indication 
for any X-ray absorption in considerable excess to the
optical extinction to \AG 
(E$_{\mathrm{B-V}} = 0.1\pm0.05$ mag; \citealt{kenyon_93};
E$_{\mathrm{B-V}} = 0.09\pm0.04$ mag; \citealt{vogel_94}),
although the derived values of the X-ray absorption may mean
some slight increase of the circumstellar absorption after 
the end of the outburst of \AG in 2015.

Finally, it is worth noting that the X-ray emission of \AG in 2017
returned to its level before the outburst of this
object in 2015. The 2017 observed flux in the 0.3 - 3 keV energy range 
of $6.21\times10^{-13}$ ergs cm$^{-2}$ s$^{-1}$ is about a factor of
2 lower than the average flux in 2015  and comparable to the value of
the pre-outburst observations in 2013 (see table 1 and fig. 2 in
\citealt{zht_16}).

\subsection{Variability}
\label{sec:xray_lc}

The \XMM observation of \AG allows to search for variability on
time-scale less than $\sim 30$~ks both in X-rays and in UV. 
To check for variability, we fitted each LC with a constant, adopting
$\chi^2$ fitting.

\begin{figure*}
\begin{center}
\includegraphics[width=2.0in, height=1.43in]{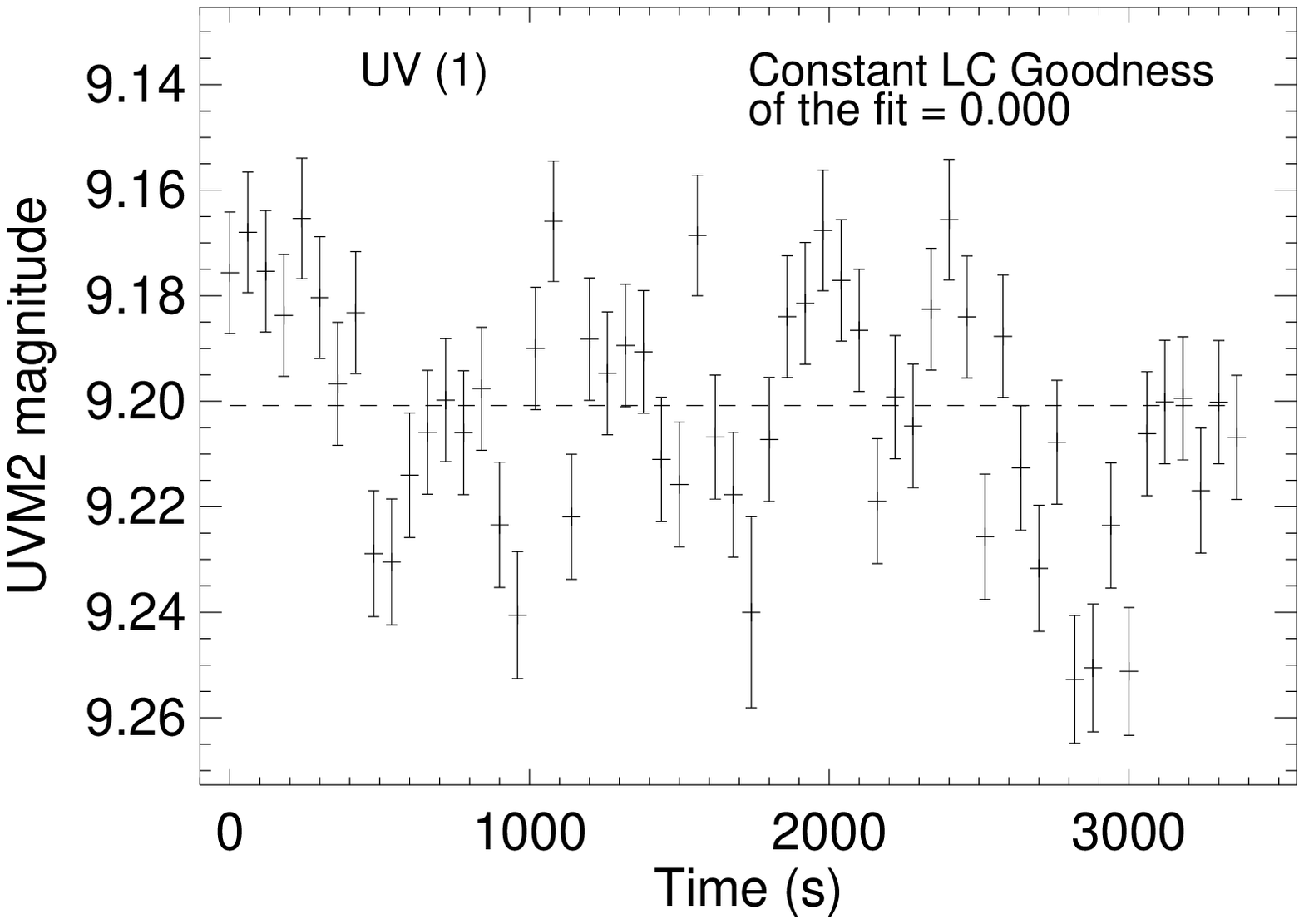}
\includegraphics[width=2.0in, height=1.43in]{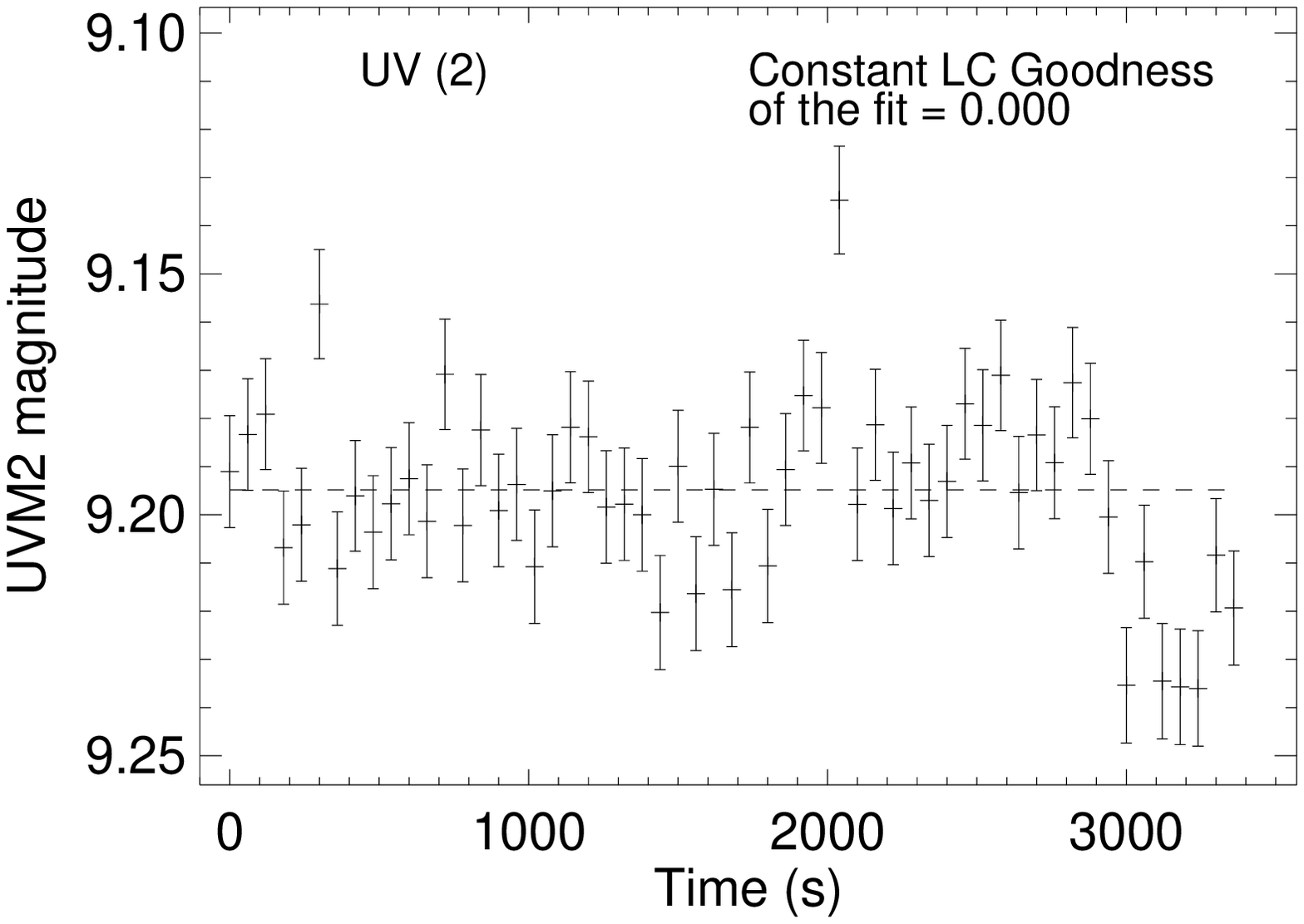}
\includegraphics[width=2.0in, height=1.43in]{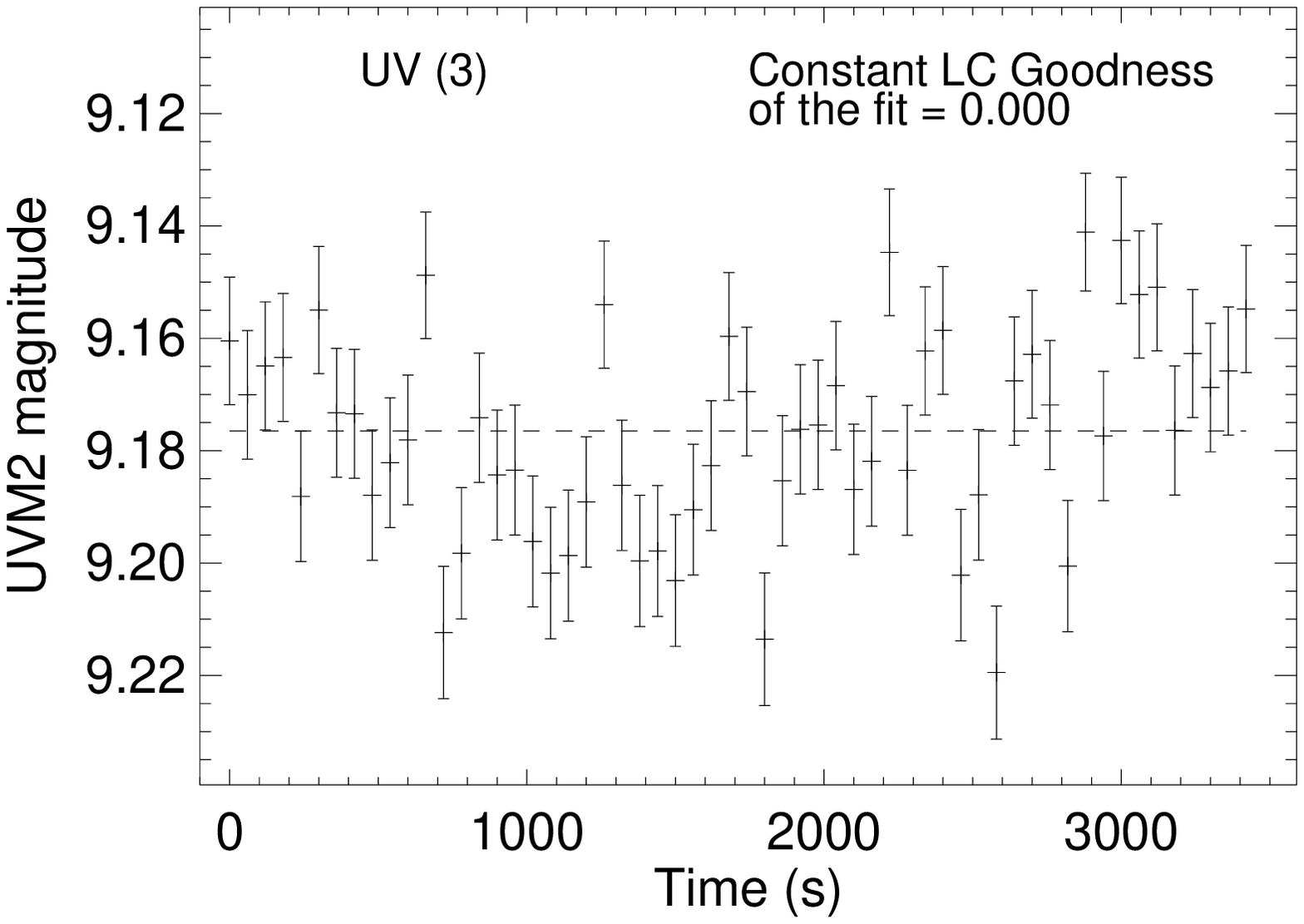}
\includegraphics[width=2.0in, height=1.43in]{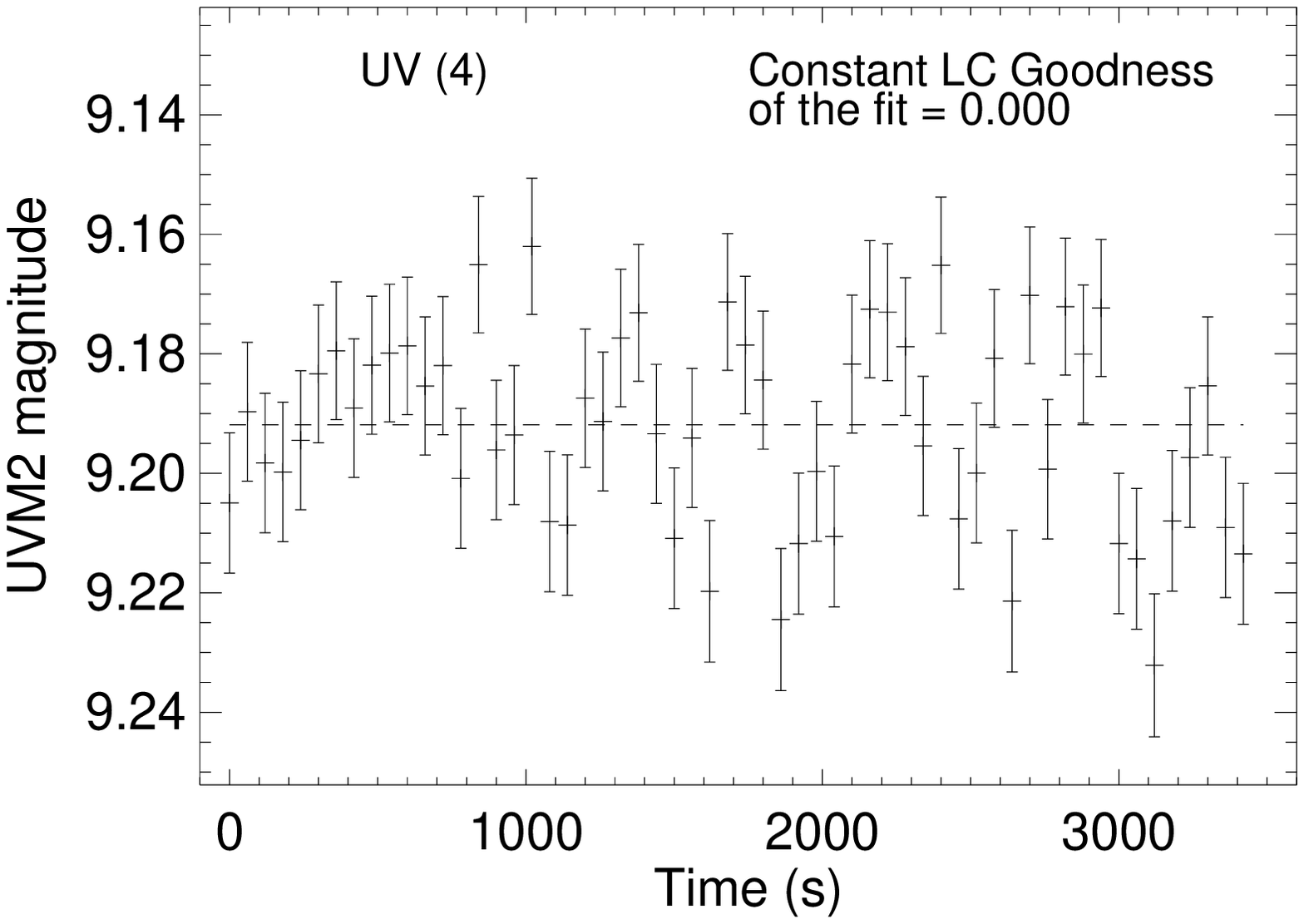}
\includegraphics[width=2.0in, height=1.43in]{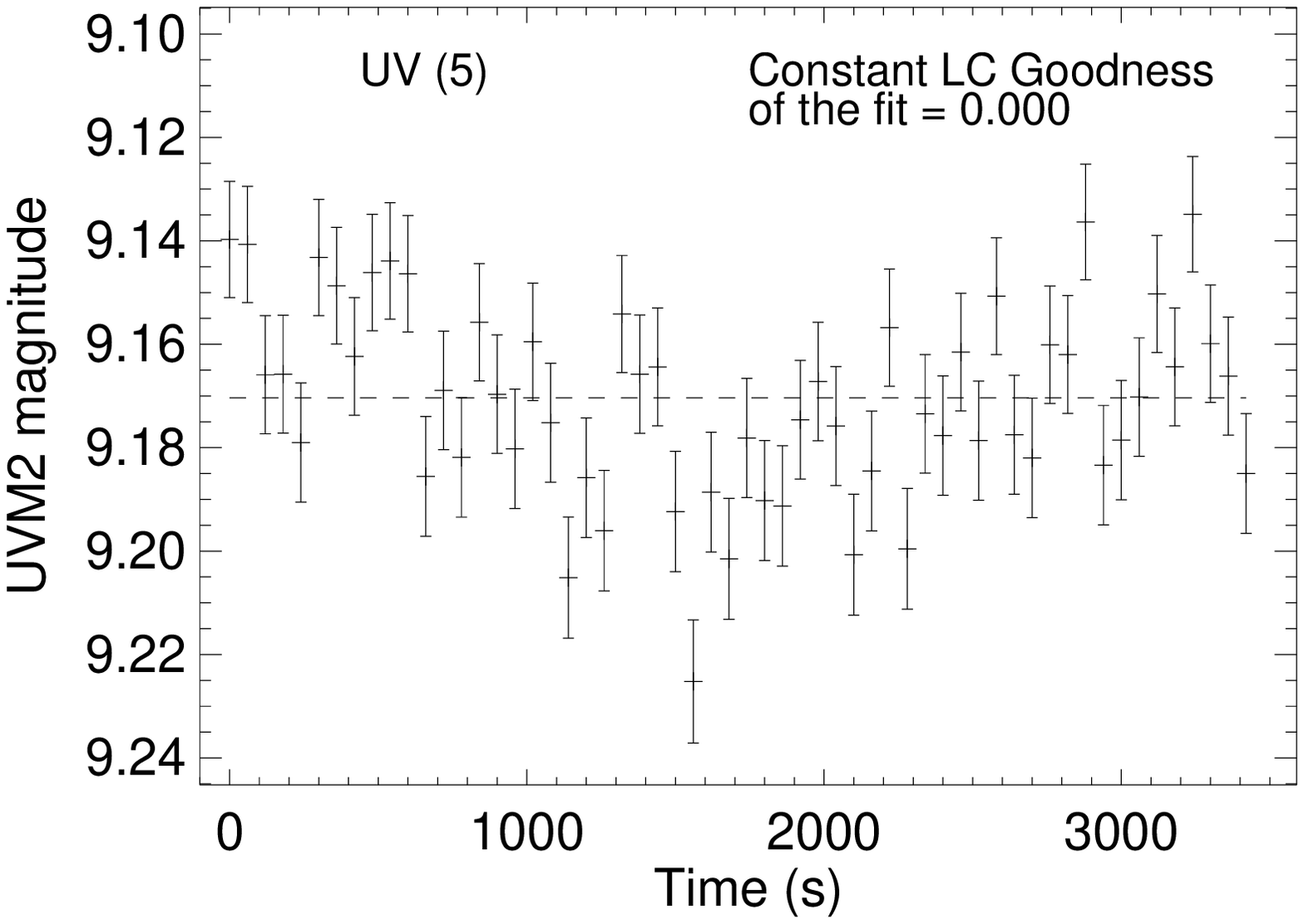}
\includegraphics[width=2.0in, height=1.43in]{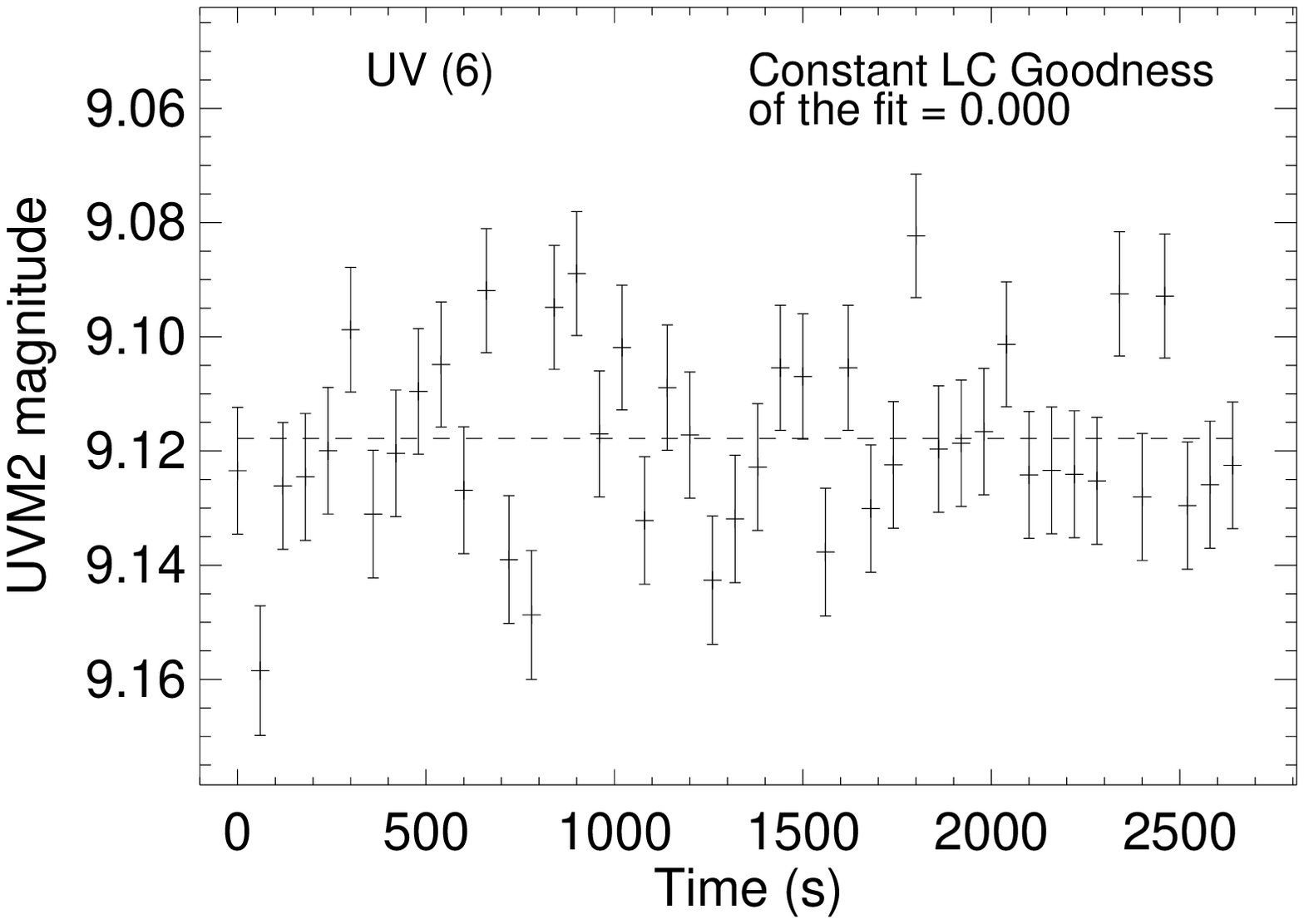}
\includegraphics[width=2.0in, height=1.43in]{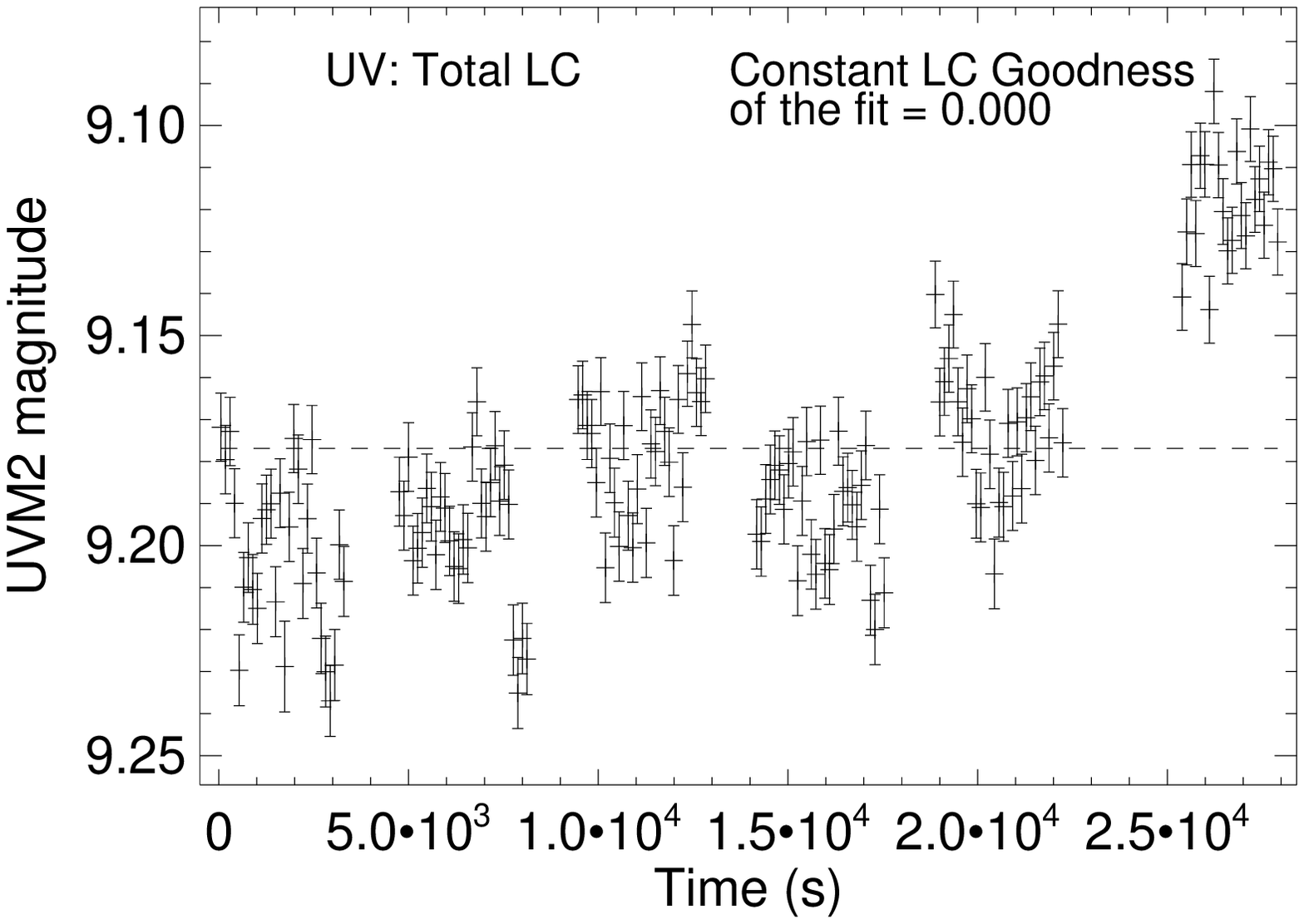}
\includegraphics[width=2.0in, height=1.43in]{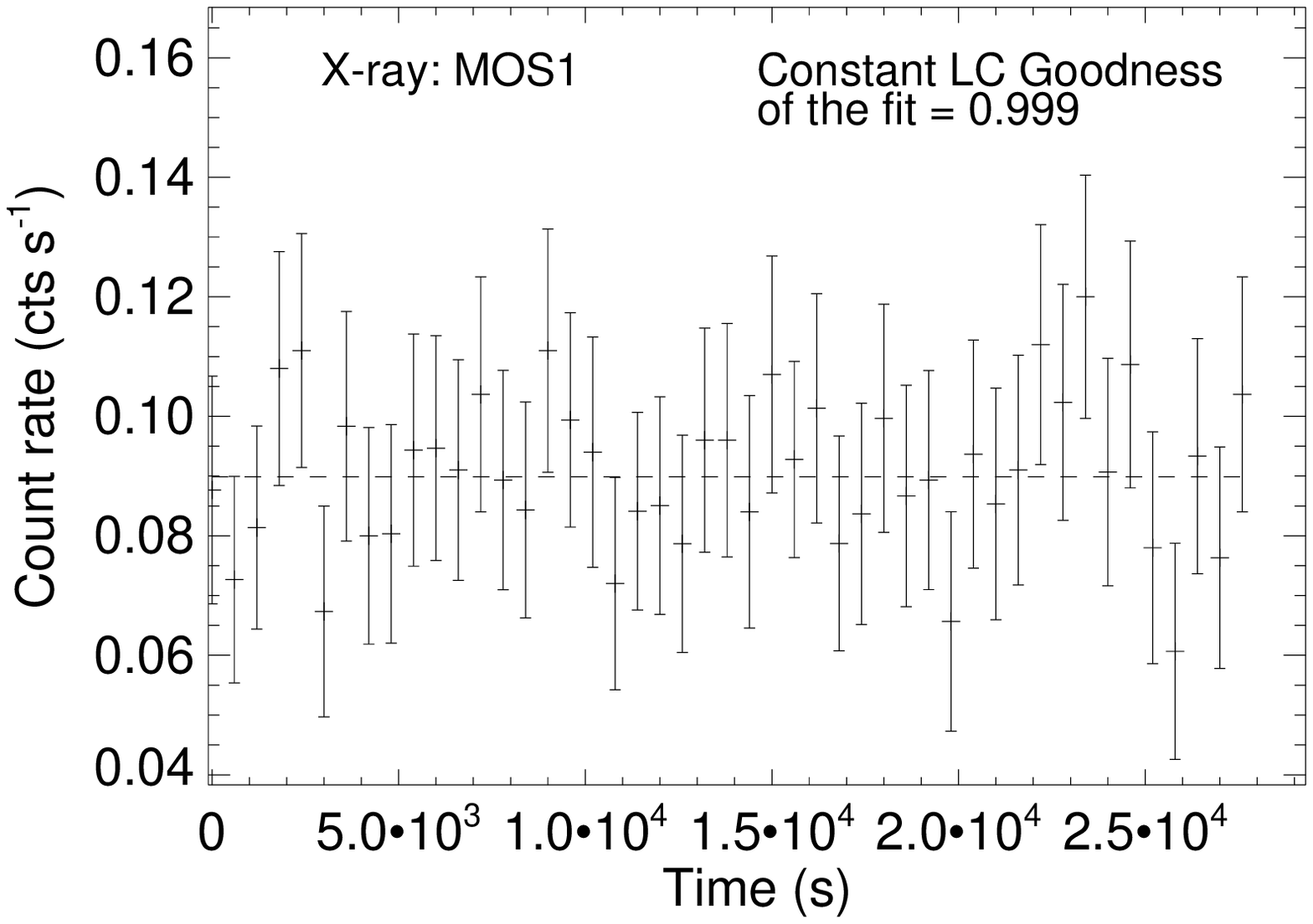}
\includegraphics[width=2.0in, height=1.43in]{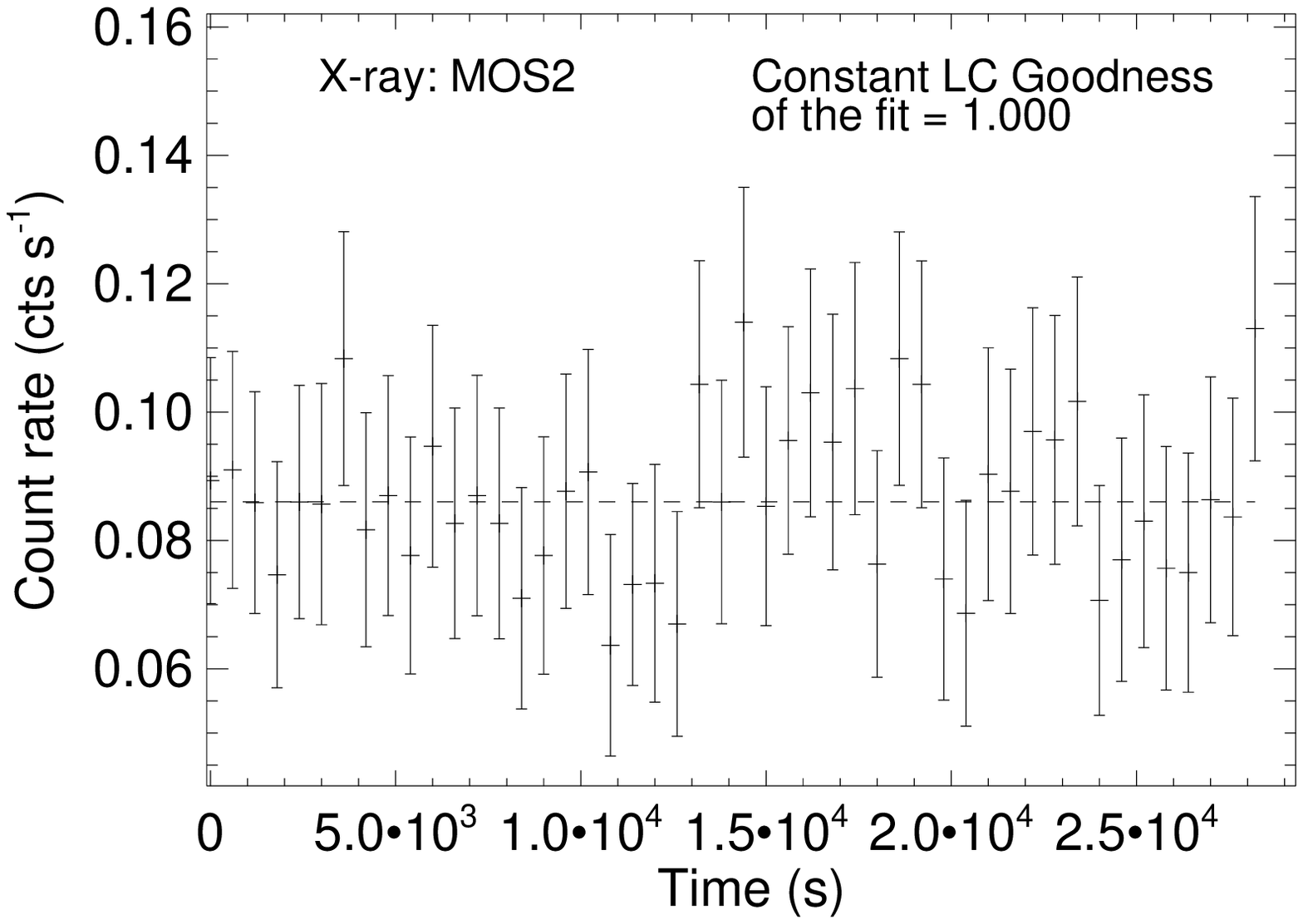}
\end{center}
\caption{The \XMM light curves (LC) of  \AGE.
{\it First and second} rows:
LCs from the six individual exposures of the \XMM OM (optical
monitor), denoted UV (1)...(6), respectively.
{\it Third} row: the total UV LC (first panel) and the X-ray LCs
(from MOS1 and MOS2 detectors).
The corresponding constant magnitude (or count rate) level is denoted
by a dashed line.
}
\label{fig:LCs}
\end{figure*}

{\it In X-rays}, we made use of the light curves from the MOS1 and
MOS2 detectors, adopting various binning. On a time-scale less than 
30 ks and time bins between 100 and 1000 s, the X-ray light curves 
were statistically consistent with a constant flux. This result is
confirmed for the LC from the pn detector: it spans only
$\sim 10$~ks due to considerable loss of the exposure due to high
X-ray background. The MOS1 and MOS2 LCs binned at 600 s
are shown in Fig.~\ref{fig:LCs}.

{\it In UV}, we made use of the LCs from all six individual exposures,
also adopting different binning between 10 (original time bin) and 
300 s. No matter the time-bin size, stochastic variability
({\it flickering}) was always present and in all the individual 
exposures. This result does not depend on the format of the light 
curve: it is valid for the LCs in flux units (counts s$^{-1}$) and in 
magnitudes.
Also, we calculated the power spectrum for each LC but no periodic 
signal was detected on time scales smaller than the time extent of the
corresponding LC.
The individual UV LCs (binned at 60 s) and the total UV LC (binned at
120 s) are given in Fig.~\ref{fig:LCs}.

Such a different behaviour of the X-ray and UV light curves likely
indicates that the X-ray and UV emissions originate from different
regions in \AG and are subject to different formation mechanisms.

\section{Discussion}
\label{sec:discussion}

From analysis of the first \AG detection with \RosatE, 
\citet{murset_95} reported soft X-ray emission from an optically thin
plasma with a temperature of a few $10^6$~ K. These authors suggested 
that the X-ray emission likely originates from colliding stellar winds 
(CSWs), that is from the interaction region of the winds from the cool 
and hot stellar components in this symbiotic binary.

During the active phase of \AG in 2015, \Swift observations 
revealed considerable X-ray variability (flickering) on time-scale of 
days as variability on shorter time-scale might have been present as 
well, which means that some other mechanism, not CSWs, was responsible 
for the X-ray emission of the active phase \citep{zht_16}.

\XMM observed \AG in 2017, that is after the end of its active phase,
and the basic characteristics of its X-ray emission were quite similar
to those from the early \Rosat observation (Section~\ref{sec:results}).
Namely, the X-ray emission of \AG is soft and its spectrum is well
matched by emission from optically thin, and temperature stratified, 
plasmas of a few $10^6$~ K.

Such similarities justify considering the 
X-ray emission from \AG in 2017 in the framework of the CSW picture in 
some more detail and in comparison with a similar consideration of the
\Rosat spectrum of \AGE. 

We note that the \Rosat spectrum considered in this study
is the same one as in \citet{zht_16}: its data reduction and
spectral extraction are described in section 2 therein. It is worth 
recalling that the \Rosat spectrum is not of that high quality as the
EPIC spectra are: it has only $\sim 400$ source counts (to compare 
with $\sim 5 000$ source counts in the pn and MOS spectra; see 
Section.~\ref{sec:data}).

\subsection{CSW model spectra}
\label{sec:csw_model}

We note that a direct comparison between the CSW model and observed X-ray 
emission of \AG was presented in \citet{zht_16}. Here, we will adopt
the same approach and we will briefly recall below the basic
characteristics of that CSW model and some changes of its as well.

The basic input parameters for the hydrodynamic model
in CSW binaries are the mass loss and velocity of the stellar winds 
of the binary components and the binary separation. 
They determine the shape and the structure of the CSW interaction 
region, that is the physical parameters (density, temperature,
velocity, emission measure) of the shocked plasma 
(\citealt{lm_90}; \citealt{luo_90}; \citealt{stevens_92}; 
\citealt{mzh_93}).

Given the values of the input parameters, some dimensionless
parameters are introduced that indicate whether or not some physical
processes are important that may affect the X-ray emission of the
interaction region. For example, the importance of the radiative
losses is indicated by either of dimensionless parameters
$\chi$ \citep{stevens_92} and $\Gamma_{ff}$ \citep{mzh_93}.
The importance of partial electron heating behind strong shocks is
given by the value of dimensionless parameter $\Gamma_{eq}$
\citep{zhsk_00}. The dimensionless parameter $\Gamma_{NEI}$
\citep{zh_07} determines the importance of the non-equilibrium 
ionization effects (NEI) for the X-ray emission from the CSW region.

And, given the results from the CSW hydrodynamic model
(the latter is based on the 2D numerical hydrodynamic model of CSW
by \citealt{lm_90}, see also \citealt{mzh_93}), we run the
corresponding CSW model in \xspec that is used to directly fit the
observed X-ray spectrum of the studied binary. We note that our CSW 
models were originally developed in the \xspec version 11.3.2
\citep{zh_07} but they
can now be used in \xspec 12.9.1, thus, allowing for using the recent
atomic data (see section 4.1 in \citealt{zh_17} for more details on 
the CSW models and fitting procedure).

For the CSW spectral models of \AGE, we adopted the same stellar wind 
and binary parameters as in \citet{zht_16}.
Namely, the mass-loss rates and wind velocities are
based on \citet{murset_95} and \citet{schmutz_96}:
$\dot{M}_{HS} = 2\times10^{-7}$\dotM, $V_{HS} = 1000$\kms,
$\dot{M}_{CS} = 4\times10^{-7}$\dotM, $V_{CS} = 20$\kms (the
notations $HS$ and $CS$ stand for the hot star and cool star,
respectively).
The binary separation of $3.22\times10^{13}$~cm comes from 
Kepler's third law for
an adopted
period of 818.2 days \citep{fekel_00} and assuming a total binary
mass of 2 solar masses.
For the distance to \AGE, we used the value of d~$=800$~ pc 
(\citealt{kenyon_93}; \citealt{kenyon_01}).
We recall that for these wind and binary parameters the
dimensionless parameters 
($\chi$, $\Gamma_{ff}$, $\Gamma_{eq}$, $\Gamma_{NEI}$)
suggest \citep{zht_16}:
(a) the CSWs in \AG (the shocked HS plasma) will be adiabatic;
(b) the temperature equalization is very fast (that is, the electron
and ion temperatures are equal in the shocked plasma);
(c) the NEI effects do not play an important role (the hot plasma is
in collisional equilibrium ionization, CIE).

For the abundances, we adopted the same values as in the case of
discrete-temperature spectral models (Section~\ref{sec:xray_spec}) and
for the spectra with good photon statistics (pn, MOS) the abundances 
of N, O, Ne, Mg, Si and Fe were allowed to vary to improve the quality 
of the fits.

\begin{figure*}
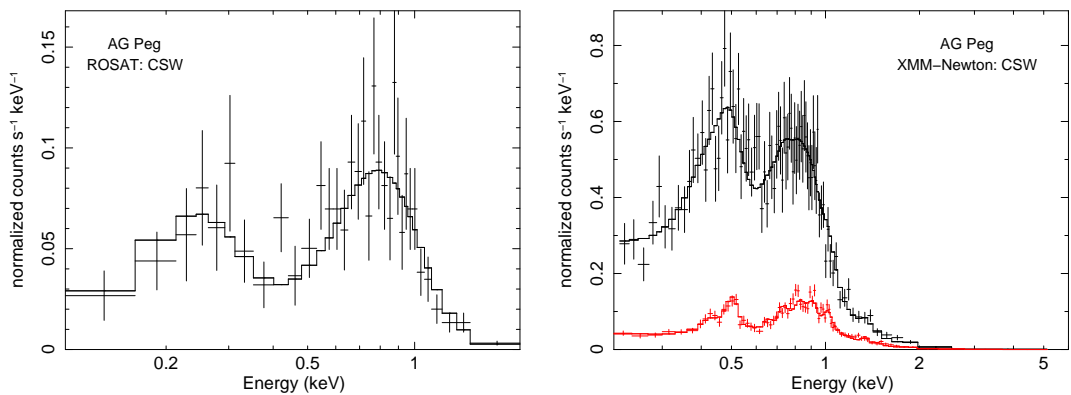

\begin{center}
\includegraphics[width=2.0in, height=2.8in, angle=-90] {fig5a.eps}
\includegraphics[width=2.0in, height=2.8in, angle=-90] {fig5b.eps}
\end{center}
\caption{The background-subtracted spectra of \AG overlaid with
the CSW model.
For \XMME, the pn (upper curve) and MOS (lower curve) spectra are
shown in black and red colour, respectively.
The spectra were re-binned to have a minimum of 50 (\XMME) and 10
(\RosatE) counts per bin.
}
\label{fig:spec_csw}
\end{figure*}

We ran a series of CSW spectral models to fit the X-ray spectra of
\AGE. This was done both for the \XMM spectra obtained in 2017 
and for the \Rosat spectrum obtained in 1993.
Table~\ref{tab:csw} and Fig.~\ref{fig:spec_csw} present some of the 
fit results.

It should be noted that the quality of the fits is good: the CSW model
matches well the shape of the observed \XMM and \Rosat spectra.
However, to get a good correspondence between the theoretical flux
(emission measure) and that observed the mass-loss rates in \AG had to
be changed with respect to their nominal values:
by a factor of 1.06 in the case of the \Rosat spectrum and with a
factor of 1.86 in the case of the \XMM spectra. 

For the \XMM spectra, the derived value of the X-ray absorption from 
the CSW model is very similar to that from the fit with the 
discrete-temperature model (Table~\ref{tab:fits}), that is it is in 
general consistent with the optical extinction to \AGE, as 
discussed in Section~\ref{sec:xray_spec}. 
The latter is true also for the X-ray absorption derived from the CSW
model fit to the \Rosat spectrum of \AGE. 

Interestingly, the observed flux in the (0.1 - 2 keV) energy range 
in the pre-outburst (\RosatE) and post-outburst (\XMME) state of \AG 
is virtually the same. The $\sim 5$\% difference is within the 
associated $1\sigma$ confidence interval on the flux: 
F$_{X} = [6.10 - 6.75]$ (\RosatE) and $[5.89 - 6.14]$ (\XMME) in units
of  $10^{-13}$ ergs cm$^{-2}$ s$^{-1}$.
Thus, using the same physical model (picture) we find again (see also
Section~\ref{sec:xray_spec}) that the pre-outburst and post-outburst 
states in \AG have very similar level of X-ray emission.

We see that the CSW model is capable of explaining
the X-ray emission of \AG in its pre-outburst state (1993 June;
\RosatE) as well as in its post-outburst state (2017 November; \XMME).
However, we have to keep in mind that there is a {\it mandatory}
requirement for the validity of the CSW picture. Namely, we must have
a solid evidence that both components in the binary system possess
stellar winds. It is well known that objects of a late spectral class
(e.g., M giants) have slow and relatively massive stellar winds but
this may not be always the case with the low-mass hot stars (i.e.,
white dwarfs). 

We just recall that for the pre-outburst state of \AG
such a solid evidence for presence of a stellar wind from the hot 
component was indeed found from analysis of high resolution UV spectra
(see \citealt{nuss_95}; \citealt{schmutz_96}). And, we
{\it do need} a similar finding for the post-outburst state of \AGE.

For the moment, we have only {\it indirect} signs that the CSW model
is a reasonable representation of the X-ray emission from \AG in its
post-outburst state.

Among others, we recall that
the dispersed spectra of \AG in 2017 show that the forbidden line in 
the He-like triplet of OVII is not suppressed 
(Section~\ref{sec:xray_overview} and Fig.~\ref{fig:lines}).
This indicates that the line emission originates from a region with 
relatively low plasma density
located far from strong sources of UV emission. That is, the X-ray 
emission is not expected to form close to or on the surface of the hot
component. So, the CSW region in \AG seems a suitable place for the
origin of the X-rays from this symbiotic binary.

Also, the strong emission lines in the dispersed spectra of \AG in
2017 show no appreciable line shifts (all are consistent with 
zero-shifts). We note that in the framework of the CSW picture,
which considers interaction of two spherically-symmetric stellar 
winds, the axis of symmetry of the CSW `cone' coincides with the line 
of centres of the binary components. And, the CSW cone
has its apex pointing to the star with the stronger wind, therefore,
the shocked plasma flow will be towards the star with the weaker wind.
Thus, 
if the orbital inclination of the binary system is not zero, the 
spectral lines from the interaction region are expected to have line
shifts that vary with the orbital phase. The maximum red- or 
blue-shifts of the lines are expected when the star with the stronger 
wind or the stellar component with the weaker wind is respectively 
`in-front' with respect to observer. Non-shifted line centres are to 
be observed when the observer's line of sight is perpendicular to the 
line of centres of the binary components. 
For orbits with low ellipticity, these are the orbital phases
near the maximum and the minimum of the radial-velocity curve of the
stellar components {\it themselves}. Using the ephemeris from
\citet{fekel_00}, we derive that the \XMM observation of \AG was 
carried out at orbital phase 0.765. Interestingly, this is exactly 
near the orbital phases with the maximum radial-velocity of the M 
giant in \AG (see fig.9 in \citealt{fekel_00}). So, no line shifts 
are to be expected for the spectral lines of the CSW region 
at that moment.

But, what about the different type of variability in the UV and X-ray
spectral domains in 2017? Namely, on short time-scales (minutes and 
hours) \AG shows noticeable stochastic variability (flickering) in UV 
and {\it no} variability in X-rays (see Section~\ref{sec:xray_lc} and
Fig.~\ref{fig:LCs}). We think this result can be understood in the 
framework of the CSW picture. Since the presumed hot-star wind is
driven by the radiation pressure of the hot component, fluctuations of
the UV radiation may cause similar variability (i.e., on about the 
same short time-scales) of the stellar wind parameters, which in turn
may result in some fluctuation of the CSW region parameters. However,
we have to keep in mind that there is a characteristic time-scale that
determines how the CSW region responds to any physical fluctuations.
This is the dynamical time ($t_d$) of the hydrodynamical problem of 
CSWs and for \AG $t_d$ is $\sim 2 -3$~days ($t_d \approx a / V_{wind}$,
where $a = 3.22\times10^{13}$~cm is the binary separation and 
$V_{wind} = 1000$\kms; see above). This value of the dynamical time
means that the CSW region in \AG will `promptly' respond to any 
changes in the stellar winds that have typical time-scale of weeks 
or months. On the other hand, all the short-time fluctuations will 
simply be smeared out and the CSW region will adopt such physical
parameters that correspond to the time-averaged values of the stellar
wind parameters. Thus, we may expect that the X-ray emission from the 
presumed CSWs in \AG to have no variability on short time-scales 
despite the appreciable UV variability (flickering) from this 
symbiotic binary.

However, we underline that if future optical/UV  observations do not find 
a solid evidence for a massive enough hot-star wind or if they find
that {\it no} hot-star wind exists now (in the
post-outburst state), then the CSW picture will face serious problems
and a different mechanism must be proposed for the origin of the X-ray
emission from \AGE.

\begin{table}
\caption{CSW Spectral Model Results
\label{tab:csw}}
\begin{center}
\begin{tabular}{lll}
\hline
\multicolumn{1}{c}{Parameter} &
          \multicolumn{1}{c}{\Rosat} & \multicolumn{1}{c}{\XMM} \\
   \\
\hline
$\chi^2$/dof  & 23/33 & 169/153
   \\
N$_{\mathrm{H}}$  & 0.43$^{+0.04}_{-0.03}$ & 0.90$^{+0.08}_{-0.08}$
   \\
N   & 8.93 & 4.62$^{+0.96}_{-0.77}$  \\
O   & 1.0 & 0.24$^{+0.05}_{-0.04}$  \\
Ne  & 1.0 & 0.60$^{+0.16}_{-0.13}$  \\
Mg  & 1.0 & 0.48$^{+0.15}_{-0.13}$  \\
Si  & 1.0 & 0.28$^{+0.14}_{-0.12}$  \\
Fe  & 1.0 & 0.26$^{+0.04}_{-0.03}$  \\
$norm$  &  1.00$^{+0.07}_{-0.07}$ & 1.00$^{+0.12}_{-0.12}$
   \\
F$_{X,1}$   &
  6.41 (13.7) & 6.08 (20.2)
   \\
F$_{X,2}$   &
  \dots & 6.38 (15.2)
   \\
\hline

\end{tabular}
\end{center}

{\it Notes.}
Results from the fit to the \Rosat
spectrum of \AG using model spectra from the CSW hydrodynamic
simulations for the case with collisional ionization equilibrium
(CIE).
Tabulated quantities are the neutral H absorption column
density (N$_{\mathrm{H}}$) in units of 10$^{21}$ cm$^{-2}$,
the normalization parameter ($norm$) and the absorbed X-ray flux
in units of $10^{-13}$ ergs cm$^{-2}$ s$^{-1}$
(F$_{X,1}$ in the 0.1 - 2 keV range;
F$_{X,2}$ in the 0.2 - 8 keV range)
followed in parentheses by the unabsorbed value. The $norm$
parameter is a dimensionless quantity that gives the ratio of observed
to theoretical fluxes. A value of $norm = 1.0$ indicates a perfect
match between the observed count rate and that predicted by the model.
The adopted abundances are with respect to
the solar abundances \citep{ag_89}.
Errors are the $1\sigma$ values from the fit.

\end{table}

\section{Conclusions}
\label{sec:conclusions}
In this work, we presented the \XMM data of \AGE, obtained after the
end of its outburst in 2015. The basic results and conclusions from 
our analysis of these data are as follows.

(i) The X-ray emission of \AG is of thermal origin as indicated by the
presence of emission lines of high-ionization species (NVII, OVII, 
OVIII, Fe XVII)  in the dispersed spectra of \AGE. A strong (not
suppressed) forbidden line in the He-like triplet of OVII is detected,  
which indicates that the X-ray emission originates from a rarefied hot 
plasma far from strong sources of ultraviolet emission.

(ii) The thermal origin of the X-ray emission and the fact that almost 
all of the source counts are at energies below 2 keV indicate
that \AG in 2017, that is in its post-outburst state, is of 
the class $\beta$ of the X-ray sources amongst the symbiotic stars.
The X-ray spectra of \AG are well matched by a two-temperature
optically-thin plasma emission ($kT_1 \sim 0.14$~keV and 
$kT_2 \sim 0.66$~keV).

(iii) No short-term (within 30 ks) X-ray variability is detected.
On the other hand, the UV emission (UVM2 optical filter with 
effective wavelength and width of 2310 \AA ~and 480 \AA, respectively)
of \AG shows stochastic variability (flickering) on time-scales of
minutes and hours. Such a different behaviour is a sign that the X-ray 
and UV emissions originate from different regions in \AG and are 
subject to different formation mechanisms.

(iv) Since the X-ray emission of symbiotic stars of the class $\beta$ 
is assumed to originate from colliding stellar winds in binary 
system (\citealt{murset_97}; \citealt{luna_13}), 
we also made use of
CSW modelling in our analysis by adopting the stellar wind and orbital 
parameters of \AG deduced from analysis in other spectral domains
(e.g., \citealt{schmutz_96}; \citealt{fekel_00}). 
The basic result is
that the CSW model spectrum can match well the shape of the \XMM 
spectra, but the mass-loss rates had to be increased by a factor of 
1.86 to get a good correspondence between the theoretical and observed 
flux values.

(v) Despite the fact that the CSW model is capable of explaining the
X-ray properties of \AG in its post-outburst state, a solid evidence
for a massive enough hot-star wind is needed from observations in
other spectral domains to proof the validity of the CSW picture.
If no such an evidence is found, then the CSW picture will face 
serious problems and a different mechanism must be proposed for the 
origin of the X-ray emission from this symbiotic binary.

\smallskip
{\bf Acknowledgements.} 
%
{\small
This research has made use of data and/or software provided by the
High Energy Astrophysics Science Archive Research Center (HEASARC),
which is a service of the Astrophysics Science Division at NASA/GSFC
and the High Energy Astrophysics Division of the Smithsonian
Astrophysical Observatory. 
This research has made use of the NASA's Astrophysics Data System, and
the SIMBAD astronomical data base, operated by CDS at Strasbourg,
France.
The authors thank an anonymous referee for 
valuable comments and suggestions.
}

\,
\vspace{-0.2cm}
\bibliographystyle{mnras}
\bibliography{agpeg} 

\begin{thebibliography}{}
\makeatletter
\relax
\def\mn@urlcharsother{\let\do\@makeother \do\$\do\&\do\#\do\^\do\_\do\%\do\~}
\def\mn@doi{\begingroup\mn@urlcharsother \@ifnextchar [ {\mn@doi@}
  {\mn@doi@[]}}
\def\mn@doi@[#1]#2{\def\@tempa{#1}\ifx\@tempa\@empty \href
  {http://dx.doi.org/#2} {doi:#2}\else \href {http://dx.doi.org/#2} {#1}\fi
  \endgroup}
\def\mn@eprint#1#2{\mn@eprint@#1:#2::\@nil}
\def\mn@eprint@arXiv#1{\href {http://arxiv.org/abs/#1} {{\tt arXiv:#1}}}
\def\mn@eprint@dblp#1{\href {http://dblp.uni-trier.de/rec/bibtex/#1.xml}
  {dblp:#1}}
\def\mn@eprint@#1:#2:#3:#4\@nil{\def\@tempa {#1}\def\@tempb {#2}\def\@tempc
  {#3}\ifx \@tempc \@empty \let \@tempc \@tempb \let \@tempb \@tempa \fi \ifx
  \@tempb \@empty \def\@tempb {arXiv}\fi \@ifundefined
  {mn@eprint@\@tempb}{\@tempb:\@tempc}{\expandafter \expandafter \csname
  mn@eprint@\@tempb\endcsname \expandafter{\@tempc}}}

\bibitem[\protect\citeauthoryear{{Anders} \& {Grevesse}}{{Anders} \&
  {Grevesse}}{1989}]{ag_89}
{Anders} E.,  {Grevesse} N.,  1989, \mn@doi [\gca]
  {10.1016/0016-7037(89)90286-X}, \href
  {http://adsabs.harvard.edu/abs/1989GeCoA..53..197A} {53, 197}

\bibitem[\protect\citeauthoryear{{Arnaud}}{{Arnaud}}{1996}]{Arnaud96}
{Arnaud} K.~A.,  1996, in {Jacoby} G.~H.,  {Barnes} J.,  eds,  Astronomical
  Society of the Pacific Conference Series Vol. 101, Astronomical Data Analysis
  Software and Systems V. p.~17

\bibitem[\protect\citeauthoryear{{Cash}}{{Cash}}{1979}]{cash_79}
{Cash} W.,  1979, \mn@doi [\apj] {10.1086/156922}, \href
  {http://adsabs.harvard.edu/abs/1979ApJ...228..939C} {228, 939}

\bibitem[\protect\citeauthoryear{{Fekel}, {Joyce}, {Hinkle}  \&
  {Skrutskie}}{{Fekel} et~al.}{2000}]{fekel_00}
{Fekel} F.~C.,  {Joyce} R.~R.,  {Hinkle} K.~H.,   {Skrutskie} M.~F.,  2000,
  \mn@doi [\aj] {10.1086/301260}, \href
  {http://adsabs.harvard.edu/abs/2000AJ....119.1375F} {119, 1375}

\bibitem[\protect\citeauthoryear{{Getman}, {Feigelson}, {Grosso},
  {McCaughrean}, {Micela}, {Broos}, {Garmire}  \& {Townsley}}{{Getman}
  et~al.}{2005}]{getman_05}
{Getman} K.~V.,  {Feigelson} E.~D.,  {Grosso} N.,  {McCaughrean} M.~J.,
  {Micela} G.,  {Broos} P.,  {Garmire} G.,   {Townsley} L.,  2005, \mn@doi
  [\apjs] {10.1086/432097}, \href
  {http://adsabs.harvard.edu/abs/2005ApJS..160..353G} {160, 353}

\bibitem[\protect\citeauthoryear{{Gorenstein}}{{Gorenstein}}{1975}]{go_75}
{Gorenstein} P.,  1975, \mn@doi [\apj] {10.1086/153579}, \href
  {http://adsabs.harvard.edu/abs/1975ApJ...198...95G} {198, 95}

\bibitem[\protect\citeauthoryear{{Kenyon}, {Mikolajewska}, {Mikolajewski},
  {Polidan}  \& {Slovak}}{{Kenyon} et~al.}{1993}]{kenyon_93}
{Kenyon} S.~J.,  {Mikolajewska} J.,  {Mikolajewski} M.,  {Polidan} R.~S.,
  {Slovak} M.~H.,  1993, \mn@doi [\aj] {10.1086/116749}, \href
  {http://adsabs.harvard.edu/abs/1993AJ....106.1573K} {106, 1573}

\bibitem[\protect\citeauthoryear{{Kenyon}, {Proga}  \& {Keyes}}{{Kenyon}
  et~al.}{2001}]{kenyon_01}
{Kenyon} S.~J.,  {Proga} D.,   {Keyes} C.~D.,  2001, \mn@doi [\aj]
  {10.1086/321107}, \href {http://adsabs.harvard.edu/abs/2001AJ....122..349K}
  {122, 349}

\bibitem[\protect\citeauthoryear{{Lebedev} \& {Myasnikov}}{{Lebedev} \&
  {Myasnikov}}{1990}]{lm_90}
{Lebedev} M.~G.,  {Myasnikov} A.~V.,  1990, Fluid Dynamics, \href
  {http://adsabs.harvard.edu/abs/1990FlDy...25..629L} {25}

\bibitem[\protect\citeauthoryear{{Luna}, {Sokoloski}, {Mukai}  \&
  {Nelson}}{{Luna} et~al.}{2013}]{luna_13}
{Luna} G.~J.~M.,  {Sokoloski} J.~L.,  {Mukai} K.,   {Nelson} T.,  2013, \mn@doi
  [\aap] {10.1051/0004-6361/201220792}, \href
  {http://adsabs.harvard.edu/abs/2013A%26A...559A...6L} {559, A6}

\bibitem[\protect\citeauthoryear{{Luna}, {Nunez}, {Sokoloski}  \&
  {Montane}}{{Luna} et~al.}{2015}]{luna_15}
{Luna} G.~J.~M.,  {Nunez} E.~N.,  {Sokoloski} L.~J.,   {Montane} B.,  2015, The
  Astronomer's Telegram, \href
  {http://adsabs.harvard.edu/abs/2015ATel.7741....1L} {7741}

\bibitem[\protect\citeauthoryear{{Luo}, {McCray}  \& {Mac Low}}{{Luo}
  et~al.}{1990}]{luo_90}
{Luo} D.,  {McCray} R.,   {Mac Low} M.-M.,  1990, \mn@doi [\apj]
  {10.1086/169263}, \href {http://adsabs.harvard.edu/abs/1990ApJ...362..267L}
  {362, 267}

\bibitem[\protect\citeauthoryear{{Munari}, {Valisa}, {Dallaporta}, {Cherini},
  {Righetti}  \& {Castellani}}{{Munari} et~al.}{2013}]{munari_13}
{Munari} U.,  {Valisa} P.,  {Dallaporta} S.,  {Cherini} G.,  {Righetti} G.~L.,
   {Castellani} F.,  2013, The Astronomer's Telegram, \href
  {http://adsabs.harvard.edu/abs/2013ATel.5258....1M} {5258}

\bibitem[\protect\citeauthoryear{{M\"{u}rset}, {Jordan}  \&
  {Walder}}{{M\"{u}rset} et~al.}{1995}]{murset_95}
{M\"{u}rset} U.,  {Jordan} S.,   {Walder} R.,  1995, \aap, \href
  {http://adsabs.harvard.edu/abs/1995A%26A...297L..87M} {297, L87}

\bibitem[\protect\citeauthoryear{{M\"{u}rset}, {Wolff}  \&
  {Jordan}}{{M\"{u}rset} et~al.}{1997}]{murset_97}
{M\"{u}rset} U.,  {Wolff} B.,   {Jordan} S.,  1997, \aap, \href
  {http://adsabs.harvard.edu/abs/1997A%26A...319..201M} {319, 201}

\bibitem[\protect\citeauthoryear{{Myasnikov} \& {Zhekov}}{{Myasnikov} \&
  {Zhekov}}{1993}]{mzh_93}
{Myasnikov} A.~V.,  {Zhekov} S.~A.,  1993, \mn@doi [\mnras]
  {10.1093/mnras/260.1.221}, \href
  {http://adsabs.harvard.edu/abs/1993MNRAS.260..221M} {260, 221}

\bibitem[\protect\citeauthoryear{{Nunez} \& {Luna}}{{Nunez} \&
  {Luna}}{2013}]{nunez_13}
{Nunez} E.~N.,  {Luna} M.~G.~J.,  2013, The Astronomer's Telegram, \href
  {http://adsabs.harvard.edu/abs/2013ATel.5324....1N} {5324}

\bibitem[\protect\citeauthoryear{{Nussbaumer}, {Schmutz}  \&
  {Vogel}}{{Nussbaumer} et~al.}{1995}]{nuss_95}
{Nussbaumer} H.,  {Schmutz} W.,   {Vogel} M.,  1995, \aap, \href
  {http://adsabs.harvard.edu/abs/1995A%26A...293L..13N} {293, L13}

\bibitem[\protect\citeauthoryear{{Ramsay}, {Luna}, {Nunez}, {Sokoloski}  \&
  {Montane}}{{Ramsay} et~al.}{2015}]{ramsay_15}
{Ramsay} G.,  {Luna} G.~J.~M.,  {Nunez} E.~N.,  {Sokoloski} L.~J.,   {Montane}
  B.,  2015, The Astronomer's Telegram, \href
  {http://adsabs.harvard.edu/abs/2015ATel.7779....1R} {7779}

\bibitem[\protect\citeauthoryear{{Ramsay}, {Sokoloski}, {Luna}  \&
  {Nu{\~n}ez}}{{Ramsay} et~al.}{2016}]{ramsay_16}
{Ramsay} G.,  {Sokoloski} J.~L.,  {Luna} G.~J.~M.,   {Nu{\~n}ez} N.~E.,  2016,
  \mn@doi [\mnras] {10.1093/mnras/stw1546}, \href
  {http://adsabs.harvard.edu/abs/2016MNRAS.461.3599R} {461, 3599}

\bibitem[\protect\citeauthoryear{{Schmutz}}{{Schmutz}}{1996}]{schmutz_96}
{Schmutz} W.,  1996, in {Benvenuti} P.,  {Macchetto} F.~D.,   {Schreier} E.~J.,
   eds, Science with the Hubble Space Telescope - II. p.~366

\bibitem[\protect\citeauthoryear{{Skopal} et~al.,}{{Skopal}
  et~al.}{2017}]{skopal_17}
{Skopal} A.,  et~al., 2017, \mn@doi [\aap] {10.1051/0004-6361/201629593}, \href
  {http://adsabs.harvard.edu/abs/2017A%26A...604A..48S} {604, A48}

\bibitem[\protect\citeauthoryear{{Stevens}, {Blondin}  \& {Pollock}}{{Stevens}
  et~al.}{1992}]{stevens_92}
{Stevens} I.~R.,  {Blondin} J.~M.,   {Pollock} A.~M.~T.,  1992, \mn@doi [\apj]
  {10.1086/171013}, \href {http://adsabs.harvard.edu/abs/1992ApJ...386..265S}
  {386, 265}

\bibitem[\protect\citeauthoryear{{Tomov}, {Stoyanov}  \& {Zamanov}}{{Tomov}
  et~al.}{2016}]{tomov_16}
{Tomov} T.~V.,  {Stoyanov} K.~A.,   {Zamanov} R.~K.,  2016, \mn@doi [\mnras]
  {10.1093/mnras/stw2012}, \href
  {http://adsabs.harvard.edu/abs/2016MNRAS.462.4435T} {462, 4435}

\bibitem[\protect\citeauthoryear{{Vogel} \& {Nussbaumer}}{{Vogel} \&
  {Nussbaumer}}{1994}]{vogel_94}
{Vogel} M.,  {Nussbaumer} H.,  1994, \aap, \href
  {http://adsabs.harvard.edu/abs/1994A%26A...284..145V} {284, 145}

\bibitem[\protect\citeauthoryear{{Vuong}, {Montmerle}, {Grosso}, {Feigelson},
  {Verstraete}  \& {Ozawa}}{{Vuong} et~al.}{2003}]{vuong_03}
{Vuong} M.~H.,  {Montmerle} T.,  {Grosso} N.,  {Feigelson} E.~D.,  {Verstraete}
  L.,   {Ozawa} H.,  2003, \mn@doi [\aap] {10.1051/0004-6361:20030942}, \href
  {http://adsabs.harvard.edu/abs/2003A%26A...408..581V} {408, 581}

\bibitem[\protect\citeauthoryear{{Zhekov}}{{Zhekov}}{2007}]{zh_07}
{Zhekov} S.~A.,  2007, \mn@doi [\mnras] {10.1111/j.1365-2966.2007.12450.x},
  \href {http://adsabs.harvard.edu/abs/2007MNRAS.382..886Z} {382, 886}

\bibitem[\protect\citeauthoryear{{Zhekov}}{{Zhekov}}{2017}]{zh_17}
{Zhekov} S.~A.,  2017, \mn@doi [\mnras] {10.1093/mnras/stx2309}, \href
  {http://adsabs.harvard.edu/abs/2017MNRAS.472.4374Z} {472, 4374}

\bibitem[\protect\citeauthoryear{{Zhekov} \& {Skinner}}{{Zhekov} \&
  {Skinner}}{2000}]{zhsk_00}
{Zhekov} S.~A.,  {Skinner} S.~L.,  2000, \mn@doi [\apj] {10.1086/309176}, \href
  {http://adsabs.harvard.edu/abs/2000ApJ...538..808Z} {538, 808}

\bibitem[\protect\citeauthoryear{{Zhekov} \& {Tomov}}{{Zhekov} \&
  {Tomov}}{2016}]{zht_16}
{Zhekov} S.~A.,  {Tomov} T.,  2016, \mn@doi [\mnras] {10.1093/mnras/stw1339},
  \href {http://adsabs.harvard.edu/abs/2016MNRAS.461..286Z} {461, 286}

\makeatother
\end{thebibliography}

\bsp    
\label{lastpage}
\end{document}